\documentclass[aps,prl,twocolumn,showpacs]{revtex4}
\usepackage[dvips]{graphicx}% Include figure files
\usepackage{epsf}
\usepackage{dcolumn}% Align table columns on decimal point
\usepackage{bm}% bold math
\usepackage{amsmath}
\usepackage{amssymb}
\usepackage{color} 
\usepackage{ulem}
\usepackage[%dvipdfmx,
colorlinks=true,
linkcolor=magenta,
urlcolor=magenta,
citecolor=magenta,
filecolor=blue,
pagecolor=blue,
linktocpage=true,
bookmarks=true,
bookmarksnumbered=true
]{hyperref}

\begin{document}

%\preprint{APS/123-QED}

\title{
Spin-orbit-induced exotic insulators in a three-orbital Hubbard model\\
 with $(t_{2g})^5$ electrons
}

\author{Toshihiro Sato$^1$}
\author{Tomonori Shirakawa$^{1,2}$}
\author{Seiji Yunoki$^{1,2,3}$}

\affiliation{
$^1$Computational Condensed Matter Physics Laboratory, RIKEN, Wako, Saitama 351-0198, Japan\\
$^2$Computational Materials Science Research Team, RIKEN Advanced Institute for Computational Science (AICS), Kobe, Hyogo 650-0047, Japan\\
$^3$Computational Quantum Matter Research Team, RIKEN Center for Emergent Matter Science (CEMS), Wako, Saitama 351-0198, Japan}

\date{\today}% It is always \today, today,

\begin{abstract}
On the basis of the multi-orbital dynamical mean field theory,
a three-orbital Hubbard model with a relativistic spin-orbit coupling (SOC) 
is studied at five electrons per site. The numerical calculations 
are performed by employing the continuous-time quantum Monte Carlo (CTQMC) 
method based on the strong coupling expansion. 
We find that appropriately choosing bases, i.e., the maximally spin-orbit-entangled bases, 
drastically improve the sign problem in the CTQMC calculations, 
which enables us to treat exactly the full Hund's coupling and pair hopping terms.
This improvement is also essential to reach at low temperatures 
for a large SOC region where the SOC most significantly affects the electronic structure. 
We show that a metal-insulator transition is induced by the SOC for fixed Coulomb interactions. 
The insulating state for smaller Coulomb interactions is antiferromagnetically ordered 
with the local effective total angular momentum $j=1/2$, in which the $j=1/2$ based band 
is essentially half-filled while the $j=3/2$ based bands are completely occupied. 
More interestingly, for larger Coulomb interactions, we find that an excitonic insulating state emerges, 
where the condensation of an electron-hole pair in the $j=1/2$ and $j=3/2$ based bands occurs. 
The origin of the excitonic insulator as well as the experimental implication %of our findings are discussed. 
is discussed. 
\end{abstract}
\pacs{71.27.+a, 71.30.+h, 75.25.Dk}

% 71.27.+a	Strongly correlated electron systems; heavy fermions
% 71.30.+h	Metal-insulator transitions and other electronic transitions
% 71.20.-b	Electron density of states and band structure of crystalline solids
% 75.25.Dk	Orbital, charge, and other orders, including coupling of these orders

\maketitle

Recent experiments have reported interesting observations for $5d$ transition metal Ir oxides 
in the layered perovskite structure such as $\rm Sr_{2}IrO_{4}$ and $\rm Ba_{2}IrO_{4}$~\cite{Randall,Cava,Shimura,Cao,Okabe}. 
In these materials, along with moderate electron correlations~\cite{Krempa,Watanabe3}, there exists a strong relativistic 
spin-orbit coupling (SOC), 
which splits $t_{2g}$ orbitals, already separated from $e_g$ orbitals due to a large crystal field, 
into the effective total angular momentum $j=1/2$ doublet and $j=3/2$ quartet orbitals in the atomic limit~\cite{Sugano}.
Since there are nominally five $5d$ electrons per Ir ion, the $j=1/2$ orbital 
is half filled while the $j=3/2$ orbitals are fully occupied. 

As opposed to simple expectation from strongly correlated 
$3d$ and $4d$ transition metal oxides~\cite{Matsuno,Wang,Kim3,Perry,Nagai}, 
the experiments have revealed that the ground state of these Ir oxides 
is a $j=1/2$ antiferromagnetic (AF) insulator~\cite{Kim1,Kim2,Ishii,Fujiyama}.
The theoretical understanding of the $j=1/2$ AF insulator has been also 
reported~\cite{Kim1,Jackeli,Jin,Watanabe1,Shirakawa,Martins,Arita,Onishi}.
Moreover, even possible unconventional superconductivity has been proposed once mobile carriers are introduced into the insulating 
state~\cite{Wang2,Watanabe2,Yang,Meng}. 
However, 
the electronic structure in multi-orbital systems 
with the competition between the electron correlations and the SOC has not been thoroughly understood.
When the SOC is significantly large, the $j=1/2$ based band is completely separated from the $j=3/2$ based bands,
and thus a single-orbital description of the $j=1/2$ based band is expected to be valid. 
On the other hand, when the SOC is small, this picture breaks down and 
the electronic structure should be largely affected not only 
by Coulomb interactions but also by the multi-orbital nature. 
Therefore, a question naturally arises: what is the ground state of multi-orbital systems with the competition between 
the electron correlations and the SOC. 
This is precisely the main issue of this paper and we will demonstrate the emergence of exotic 
insulators in a three-orbital Hubbard model with the SOC.

Here, on the basis of the multi-orbital dynamical mean field theory (DMFT)~\cite{MODMFT}, we numerically study 
a three-orbital Hubbard model with the SOC at five electrons per site, corresponding to $(t_{2g})^5$ electronic configuration. 
The continuous-time quantum Monte Carlo (CTQMC) method based on the strong coupling expansion is employed as 
a multi-orbital impurity solver~\cite{CTQMC}. 
We find that the sign problem is significantly improved by appropriately choosing the maximally spin-orbit-entangled bases, 
which allows us to treat exactly the full Hund's coupling and pair hopping terms. 
This improvement is crucial also for the calculations in a low temperature and a strong SOC regions where the metal-insulator transition 
(MIT) occurs. 
We find that for fixed Coulomb interactions the MIT is induced by increasing the SOC. 
The insulating phases include the $j=1/2$ antiferromagnetically ordered phase as well as a multi-orbital AF insulating (AFI) phase.
In addition, we find an excitonic insulating (EXI) phase,
where an electron-hole pair in the $j=1/2$ and $j=3/2$ based bands are condensed.

The three-orbital Hubbard model studied here is described by the following Hamiltonian: 
$H=H_0+H_{\rm I}$, where 
$
H_0=\sum_{\langle i,i^{\prime} \rangle}\sum_{\gamma,\sigma}t^{\gamma}c_{i\gamma\sigma}^{\dagger} c_{i^{\prime}\gamma\sigma}
-\mu \sum_{i,\gamma,\sigma} n_{i\sigma}^\gamma\nonumber 
+\lambda \sum_{i,\gamma,\delta,\sigma,\sigma'}\langle \gamma|\mathbf L_{i}|\delta \rangle \cdot \langle \sigma|\mathbf S_{i}|\sigma' \rangle c_{i\gamma\sigma}^{\dagger} c_{i\delta\sigma'}
$ 
represents the non-interacting part of the model and
$
H_{\rm I}=U\sum_{i,\gamma}n_{i\uparrow}^\gamma n_{i\downarrow}^\gamma+\frac{U'-J}{2}\sum_{i,\gamma\neq\delta,\sigma}n_{i\sigma}^\gamma n_{i\sigma}^\delta 
+\frac{U'}{2}\sum_{i,\gamma\neq\delta,\sigma}n_{i\sigma}^\gamma n_{i\bar{\sigma}}^\delta
-J\sum_{i,\gamma\neq\delta}c_{i\gamma\uparrow}^{\dagger} c_{i\gamma\downarrow} c_{i\delta\downarrow}^{\dagger}  c_{i\delta\uparrow} 
+J'\sum_{i,\gamma\neq\delta}c_{i\gamma\uparrow}^{\dagger}  c_{i\gamma\downarrow}^{\dagger}  c_{i\delta\downarrow} c_{i\delta\uparrow}
$
describes the local Coulomb interactions. 
Here, 
$t^{\gamma}$ sets the nearest-neighbor hopping amplitude for
$t_{2g}$ orbitals $\gamma=(d_{yz}, d_{zx}, d_{xy})$ on the Bethe lattice with coordination number $Z$~\cite{Eckstein}. 
We adopt a semielliptic density of states for $Z\to \infty$ 
assuming the same bandwidth for the three orbitals ($t^{\gamma}=t/\sqrt{Z}$), 
where the DMFT is exact~\cite{Eckstein,Metzner}. 
$c_{i\gamma\sigma}^{\dagger}$ ($c_{i\gamma\sigma}$) is an electron creation (annihilation) operator with spin 
$\sigma\,(=\uparrow,\downarrow)$ and orbital $\gamma$ at site $i$ 
and $n_{i\sigma}^{\gamma}=c_{i\gamma\sigma}^{\dagger} c_{i\gamma\sigma}$. 
$\lambda$ is the SOC and $\mathbf L_{i}$ ($\mathbf S_{i}$) is the orbital (spin) angular momentum operator at site $i$. 
The chemical potential $\mu$ is tuned to be at five electrons per site.  
$\bar\sigma$ denotes the opposite spin of $\sigma$. 
$H_{\rm I}$ includes the intra (inter) orbital Coulomb interaction $U$ ($U'$), the Hund's coupling ($J$), 
and the pair hopping ($J'$). We set $U=U'+2J$ and $J=J'=0.15U$~\cite{Kanamori}. 
Employing the CTQMC method as an impurity solver~\cite{CTQMC} to calculate the imaginary-time 
Green's functions at the impurity site, 
$G_{\gamma,\sigma}^{\delta,\sigma'}(i,\tau)\equiv-\langle T_{\tau}c_{i \gamma \sigma}(\tau)c_{i \delta \sigma'}^\dagger(0)\rangle $,
we can solve numerically exactly the model.  
In what follows, $U$, $\lambda$, temperature $T$, and frequency $\omega$ are in units of $t$. 
We also omit the site index in the Green's function, $G_{\gamma,\sigma}^{\delta,\sigma'}(\tau) = G_{\gamma,\sigma}^{\delta,\sigma'} (i,\tau)$, 
unless $G_{\gamma,\sigma}^{\delta,\sigma'}(i,\tau)$ is site dependent.

\begin{figure}[b]
\centering
\vspace{-0cm}
\centerline{\includegraphics[width=0.9\hsize]{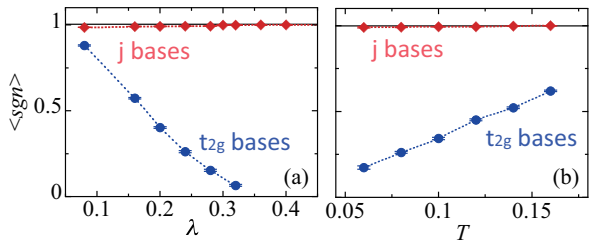}}
\vspace{-0.15cm}
\caption{(Color online) (a) $\lambda$ dependence of the average sign $\langle sgn \rangle$ for $U=8$ at $T=0.08$. 
(b) $T$ dependence of $\langle sgn \rangle$ for $U=8$ at $\lambda=0.24$. 
For both $t_{2g}$ and $j$ bases in the CTQMC calculations, the paramagnetic and orbital disordered solutions are assumed. 
}  
\label{fig:sign}
\end{figure}

Let us first examine the numerical accuracy of the CTQMC calculations.  
Generally, a negative sign problem is one of the most serious issues in QMC calculations. 
Specially, when QMC methods are employed for the multi-orbital DMFT calculations, 
the sign problem seriously prevents us from simulating at low temperatures, which often enforces 
to approximate the Hund's coupling and pair hopping terms~\cite{problem}. Indeed, for our model, we find that 
the sign problem becomes destructively serious at low temperatures specially when the SOC is large (see Fig.~\ref{fig:sign}). 

It is now important to recall that the sign problem in QMC calculations is a basis dependent problem and it can be improved 
by appropriately choosing bases~\cite{Sato}. 
Although we encounter the serious sign problem when the original $t_{2g}$ bases ($c_{i\gamma\sigma}$) 
are used, we find that the maximally spin-orbit-entangled $j$ bases ($a_{ijm}$) improve the sign problem 
significantly. These $j$ bases are the eigenstates of $H_0$ in the atomic limit and are related to the  $t_{2g}$ bases, 
\begin{eqnarray}
\left(
    \begin{array}{c}
      a_{i\frac{1}{2}\frac{s}{2}} \\
      a_{i\frac{3}{2}\frac{s}{2}} \\
      a_{i\frac{3}{2}\frac{-3s}{2}}
    \end{array}
  \right)
=\frac{1}{\sqrt{6}}
\left(\begin{array}{ccc}
\sqrt{2} & -i\sqrt{2}s & \sqrt{2}s \\
s & -i & -2 \\
 -\sqrt{3}s & -i\sqrt{3} & 0 \\
\end{array} 
\right)
\left(
    \begin{array}{c}
      c_{id_{yz}\bar\sigma} \\
      c_{id_{zx}\bar\sigma} \\
      c_{id_{xy}\sigma}
    \end{array}
  \right),
  \label{eq:a}
\end{eqnarray}
where $s=1(-1)$ for $\sigma=\uparrow(\downarrow)$. 
In the $j$ bases representation, the number of off-diagonal elements in $G_{\gamma,\sigma}^{\delta,\sigma'}(\tau)$ is reduced 
and as a result the sign problem can be alleviated.

Figure~\ref{fig:sign} shows typical results of the average sign $\langle sgn \rangle$ 
(see Ref.~\onlinecite{Sato} for the definition) for $U=8$ 
calculated for the paramagnetic and orbital 
disordered solutions. The $\lambda$ dependence of $\langle sgn \rangle$ at $T=0.08$ in Fig.~\ref{fig:sign}(a) clearly 
shows that $\langle sgn \rangle\approx1$ for the $j$ bases, much larger than 
$\langle sgn \rangle$ for the $t_{2g}$ bases particularly for large $\lambda$. 
Similarly, the $T$ dependence of $\langle sgn \rangle$ at $\lambda=0.24$ 
in Fig.~\ref{fig:sign}(b) demonstrates 
the remarkable improvement of $\langle sgn \rangle$ for low $T$ when the $j$ bases are used. 
These improvements of $\langle sgn \rangle$ guarantee higher accuracy of our CTQMC calculations
with less computational cost in a wide range of $\lambda$ even at very low $T$.

\begin{figure}
\centering
\vspace{0cm}
\centerline{\includegraphics[width=1\hsize]{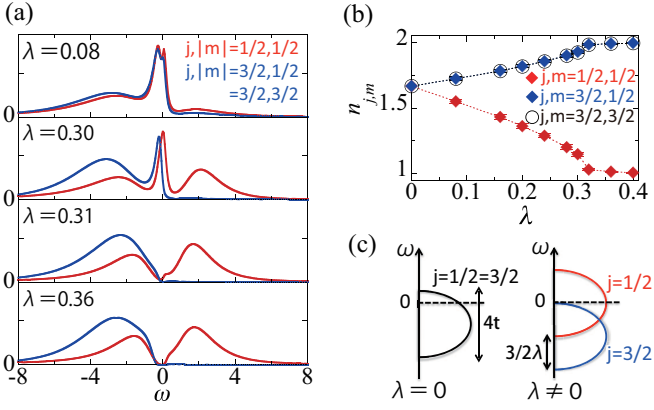}
\vspace{-0.15cm}
}
\caption{(Color online) 
(a) Single-particle excitation spectrum $A_{j,m}(\omega)$ 
and (b) electron density $n_{j,m}=\sum_{m'=\pm m}\langle a_{ijm'}^{\dag} a_{ijm'} \rangle$ 
for $U=8$ at $T=0.06$ with varying $\lambda$. 
Paramagnetic and orbital disorder states are assumed. 
(c) Schematic density of states for the non-interacting limit with $\lambda=0$ and $\lambda\ne0$. 
Fermi level is at $\omega=0$. The $j=3/2$ based bands are completely filled when $\lambda\ge 4/3$. 
}  
\vspace{-0.5cm}
\label{fig:dos-pm}
\end{figure}

First, we shall briefly examine the MIT induced by the SOC in paramagnetic and orbital disordered states. 
A typical example of the evolution of single-particle excitation spectrum 
$A_{j,m}(\omega)=-1/\pi {\rm Im}G_{j,m}^{j,m}(\omega+i0^+)$ in the real frequency $\omega$ 
with increasing $\lambda$ 
is shown in Fig.~\ref{fig:dos-pm}(a). Here, $0^+$ is positive infinitesimal. 
The maximum entropy method is employed to calculate $A_{j,m}(\omega)$
from $G_{j,m}^{j,m}(\tau)$~\cite{MEM}. 
As shown in Fig.~\ref{fig:dos-pm}(a) for $U=8$, the SOC induces the transition from 
the metallic to insulating state at $\lambda\sim0.31$,
where the quasi-particle peak near the Fermi level vanishes 
and the single-particle excitation gap starts to open. 
The insulating state for $\lambda\agt0.31$ is the $j=1/2$ Mott insulator, where the $j=1/2$ based band is half-filled while 
the $j=3/2$ based bands are fully occupied [see Fig.~\ref{fig:dos-pm}(b)]. 
Figure~\ref{fig:dos-pm}(b) also indicates that the $j=1/2$ based band, which is degenerate with the $j=3/2$ based bands 
at $\lambda=0$, gradually separates form the $j=3/2$ based bands with increasing $\lambda$ [see also Fig.~\ref{fig:dos-pm}(c)]. 
This is essential for the SOC induced Mott transition because the critical $U$ for the MIT decreases as the orbital 
degeneracy is degraded~\cite{paraMott}. 

Now, we shall consider possible ordered states.
To this end, we introduce magnetic order parameters,
$
M_{j,m}(l)=\frac{1}{2}\sum_{m^{\prime}=\pm m}sign(m^{\prime})\langle a_{ljm^{\prime}}^{\dagger} a_{ljm^{\prime}}\rangle, 
$
where $l\,(=A,B)$ indicates two sublattices~\cite{note2} 
and $a_{ljm}$ is defined in Eq.~(\ref{eq:a}) with site $i$ being on sublattice $l$. 
In addition, we investigate excitonic orders formed by an electron-hole pair in different $(j,m)$ bands with excitonic order 
parameters, 
$
\Delta_{j,m}^{j',m'}(l)=\langle a_{ljm}^{\dagger} a_{lj'm'}\rangle,
$
where $(j,m)\neq(j',m')$. 
We also calculate the electron density 
$n_{j,m}(l)=\sum_{m'=\pm m}\langle a_{ljm'}^{\dagger} a_{ljm'}\rangle$, 
$A_{j,m}(l,\omega)$, 
and the anomalous excitation spectrum 
$F_{j,m}^{j',m'}(l,\omega)$$=$$-1/\pi {\rm Im}G_{j,m}^{j',m'}(l,\omega+i0^+)$~\cite{note3,ANG}.
Since we have found that these order parameters satisfy
$M_{j,m}(A)=-M_{j,m}(B)$
and $\Delta_{j,m}^{j',m'}(A)=\Delta_{j,-m}^{j',-m'}(B)$, as well as $n_{j,m}(A)=n_{j,m}(B)$, 
$A_{j,m}(A,\omega)=A_{j,-m}(B,\omega)$ and 
$F_{j,m}^{j',m'}(A,\omega)=F_{j,-m}^{j',-m}(B,\omega)$, 
we will drop the sublattice index $l$ in these quantities hereafter.

\begin{figure}
\centering
\vspace{0cm}
\centerline{\includegraphics[width=1\hsize]{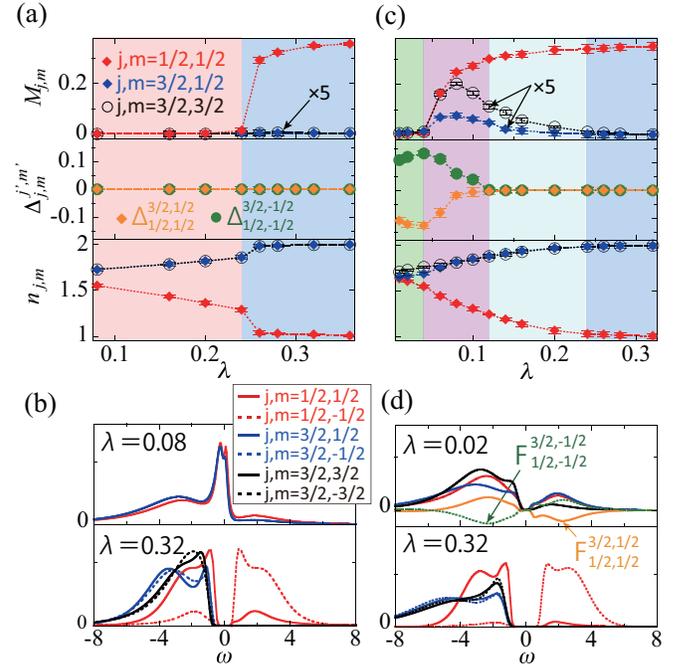}
\vspace{-0.15cm}
}
\caption{(Color online) (a) and (c): 
$\lambda$ dependence of staggered magnetization $M_{j,m}$ (top), excitonic order parameter $\Delta_{j,m}^{j,'m'}$ (center),
and electron density $n_{j,m}$ (bottom). 
(b) and (d): Single-particle excitations spectrum $A_{j,m}(\omega)$ for two $\lambda$'s. 
Other parameters used are $T=0.06$ and $U=8$ (a,b) and $9.25$ (c,d). Anomalous excitation spectrum 
$F_{1/2,\pm1/2}^{3/2,\pm1/2}(\omega)$
is also shown in (d). 
}  
\vspace{-0.5cm}
\label{fig:op-dos}
\end{figure}

Let us first explore a case with $U=8$. 
Figure~\ref{fig:op-dos} (a) shows the $\lambda$ dependence of $M_{j,m}$, $\Delta_{j,m}^{j',m'}$, and $n_{j,m}$.
First of all, it is apparent that there is no excitonic order for all values of $\lambda$. 
Second, for $\lambda \geq 0.25$, only $M_{1/2,1/2}$ is finite with $n_{1/2,1/2} \approx 1$ 
and $n_{3/2,1/2}=n_{3/2,3/2}\approx2$, revealing the $j=1/2$ AFI state. 
Third, for $\lambda \leq 0.25$, the magnetic order disappears and
$n_{3/2,1/2}=n_{3/2,3/2}$ starts to decrease from two with decreasing $\lambda$, 
implying the breakdown of the single-orbital description of the half-filled $j=1/2$ based band.

Figure~\ref{fig:op-dos}(b) shows $A_{j,m}(\omega)$ for two $\lambda$'s at $U=8$. 
Similar to the paramagnetic cases (see Fig.~\ref{fig:dos-pm}), for large $\lambda=0.32$ in the AFI phase, a finite gap is clearly 
open in the $j=1/2$ based band while the $j=3/2$ based bands are fully occupied (i.e., band insulators). 
However, notice that $A_{1/2,1/2}(\omega)$ [$A_{1/2,-1/2}(\omega)$] has larger (smaller) weight in the occupied states than in 
the unoccupied states for sublattice $A$ because of the AF order. 
For smaller $\lambda=0.08$, besides broad structures around $\omega \sim \pm U/2$, 
the sharp quasi-particle peak appears around the Fermi level for both $j$ based bands, 
indicating a metallic state with strong correlations.

Next, we shall examine a case with larger $U=9.25$. 
As shown in Fig.~\ref{fig:op-dos}(c), the $j=1/2$ AFI phase is found with no exitonic order for $\lambda\geq0.25$, the same phase 
discussed above for smaller $U$. 
$A_{j,m}(\omega)$ shown in Fig.~\ref{fig:op-dos}(d) clearly exhibits a finite gap only 
in the $j=1/2$ based band while the $j=3/2$ based bands are fully occupied. 
The difference appears as compared with the case with smaller $U$ when $\lambda$ decreases. 
For $0.12\leq\lambda\leq$$0.25$, all three $M_{j,m}$'s are now finite
and $n_{1/2,1/2}$ ($n_{3/2,1/2}$ and $n_{3/2,3/2}$) increases (decrease) from one (two) with decreasing $\lambda$ 
[seen Fig.~\ref{fig:op-dos}(c)]. 
We further confirm a finite gap in $A_{j,m}(\omega)$ for both $j$ based bands. 
These results indicate that this phase is a AFI state but apparently breaks the single-orbital description
of the half-filled $j=1/2$ based band. 

More remarkably, for $\lambda\leq0.12$, we find that the excitonic order parameter 
$\Delta_{1/2,\pm1/2}^{3/2,\pm1/2}$ is finite with non zero AF order [see Fig.~\ref{fig:op-dos}(c)]. 
Although the magnetic order eventually disappears for $\lambda\leq0.04$, $\Delta_{1/2,\pm1/2}^{3/2,\pm1/2}$ 
remains finite. 
The most important feature in this phase is that $\Delta_{1/2,1/2}^{3/2,1/2}+\Delta_{1/2,-1/2}^{3/2,-1/2}=0$ but 
$\Delta_{1/2,1/2}^{3/2,1/2}-\Delta_{1/2,-1/2}^{3/2,-1/2}$ is finite~\cite{note7} and staggered 
between the two sublattices, indicating that the excitonic order is accompanied by translational symmetry breaking. 
It is also noticed in Fig.~\ref{fig:op-dos}(c) that $n_{3/2,1/2}\ne n_{3/2,3/2}$ in this phase because of the presence of 
non-zero excitonic order. 
Moreover, we find that a single-particle excitation gap is always finite in both $j$ based bands, 
including the phase where only the excitonic order parameter is non zero, as shown in Fig.~\ref{fig:op-dos}(d). 
In addition, in the EXI phase,
$F_{1/2,\pm1/2}^{3/2,\pm1/2}(\omega)$ are found to be non zero. 
This is a strong evidence for the existence of a stable EXI state formed by an electron-hole pair 
between the $j=1/2$ and $j=3/2$ based bands with $m=\pm1/2$. In the limit of $\lambda=0$, this EXI state is replaced 
by the orbital ordered insulating state in the $t_{2g}$ orbitals reported previously~\cite{OS-3}. 

\begin{figure}
\begin{center}
\includegraphics[width=1\hsize]{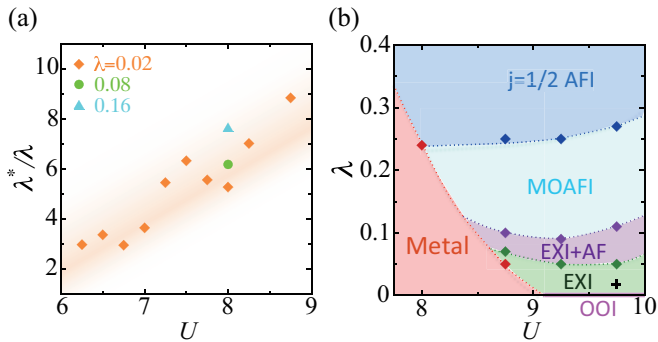}
\caption{(Color online)
(a) $U$ dependence of the effective SOC $\lambda^*$ (see the text for definition) 
for $\lambda=0.02$ at $T=0.06$. 
For comparison, the results for $\lambda=0.08$ and $0.16$ are also shown.
(b) $U$-$\lambda$ phase diagram at $T=0.06$. 
(MO)AFI, EXI, and OOI stand for (multi-orbital) antiferromagnetic insulating, excitonic insulating, 
and orbital ordered insulating phases, respectively. 
A plus mark indicates a set of $U$ and $\lambda$ where the effect of a tetragonal crystal field is examined.
}  
\label{fig:efram}
\vspace{-0.5cm}
\end{center}
\end{figure}

We shall now argue that there are two important ingredients for the emergence of the EXI state. 
First, $\lambda$ is strongly renormalized by the Coulomb interactions and 
the renormalized $\lambda$ becomes sizably large even for small $\lambda$ in the metallic phase at the vicinity of the EXI phase.  
To show this, we calculate the first moment of $A_{j,m}(\omega)$, 
defined as $W_{j,m}=\int {\rm d}\omega\, \omega A_{j,m}(\omega)$, 
and evaluate the effective SOC $\lambda^*=W_{1/2,1/2}-W_{3/2,1/2}$~\cite{note4}. 
The $U$ dependence of $\lambda^*$ for the metallic phase is shown in Fig.~\ref{fig:efram}(a). 
Indeed, $\lambda^*/\lambda$ is strongly renormalized with increasing $U$ and can be significantly large, 
as large as $\sim9$ at $U=8.75$ near the
transition to the excitonic insulator.

The other key factors are the Hund's coupling $J$ and the pair-hopping $J'$. 
We have performed the static mean-filed analysis and found that
the EXI state emerges only for a small $\lambda$ region with non zero magnetic order. 
We have also found that the excitonic order is never stabilized when $J=J'=0$. 
This implies that the particle number fluctuations are essential for the EXI state since the particle number of each $j=1/2$ and $j=3/2$ 
based bands is conserved separately when $J=J'=0$. 
These results suggest that not only the SOC but also the full Hund's coupling and the pair hopping play the important role for 
stabilizing the EXI state. The similar tendency is found in the DMFT calculations.

Finally, we have also examined the stability of the EXI state [see Fig.~\ref{fig:efram}(b)] in the case 
where the degeneracy of $t_{2g}$ orbitals 
is lifted due to a tetragonal crystal field, causing a finite splitting $\delta$ 
between the $d_{xy}$ orbital and the other two orbitals. 
We have considered both magnetic and excitonic orders in the DMFT calculations~\cite{note5}. 
Although the crystal field splitting reduces the exciton order parameter as compared to the one at $\delta=0$, we have found 
that the EXI state is robust against the crystal field splitting as long as $|\delta|$ is smaller than the single-particle excitation 
gap at $\delta=0$.

In summary, we have studied the three-orbital Hubbard model with the SOC using the multi-orbital DMFT. We have 
employed the CTQMC method based on the strong coupling expansion and found  
that the sign problem is significantly improved by using the maximally 
spin-orbit-entangled $j$ bases. The improvement is essential to treat exactly the full Hund's coupling and pair hopping terms and 
also to reach at low enough temperatures for large SOC. 

We have applied this method to determine the finite temperature phase diagram on the $U$-$\lambda$ plane at five electron per site. 
The $U$-$\lambda$ phase diagram at the lowest temperature $T=0.06$ is summarized in Fig.~\ref{fig:efram}(b). 
For small $U ( \alt 8)$, we have demonstrated that the SOC induces the transition from the metallic to the $j=1/2$ AFI state, 
in good agreement with the previous numerical study~\cite{Watanabe1}. 
For large $U (\agt 8)$, the multi-orbital AFI phase appears for intermediate values of $\lambda$, 
where the single-orbital description breaks down. 
Moreover, we have found that further overlap of the non-interacting $j=1/2$ and $j=3/2$ based bands favors the EXI phase, 
either with or without magnetic order, which dominates the phase diagram for small $\lambda\, (\alt0.1)$. 
The EXI state is formed by an electron-hole pair in the $j=1/2$ and $j=3/2$ based bands with the same $m=\pm1/2$. 
We have argued that the strong renormalization of the SOC near the MIT as well as  the full Hund's 
coupling and pair hopping are essential for the emergence of the EXI state. 

Many experimental and theoretical studies support that the $5d$ transition metal Ir oxides such as $\rm Sr_{2}IrO_{4}$ and 
$\rm Ba_{2}IrO_{4}$ are the $j=1/2$ AF insulators~\cite{Kim1,Kim2,Ishii,Fujiyama,Jackeli,Jin,Watanabe1,Shirakawa,Martins,Arita,Onishi}.
These materials are characterized with the large SOC and moderate Coulomb interactions,
thus in good qualitative accordance with our phase diagram. 
Possible materials for the excitonic insulator found here 
should be characterized with i) larger Coulomb interactions, ii) relatively small SOC, 
and iii) the low spin configuration 
with no $e_g$ orbitals involved. 
For $4d$ electron systems, relatively large Coulomb interactions and relatively small SOC are expected,
while $3d$ electron systems exhibit large Coulomb interactions and very small SOC with the high or intermediate spin configuration. 
 We expect that the ideal systems for the stable excitonic insulator would be 
somewhere between 
$3d$ and $4d$ electron systems. 
This can be achieved in experiments by, for example, $3d$ and $4d$ transition metal intermixed oxides or $4d$ transition metal oxide thin 
films grown on a compressive substrate with large distortion of oxygen octahedra. 
Further experimental studies in this direction is highly desired. 

%\section*{Acknowledgments}
The authors are grateful to K. Seki and H. Watanabe for valuable discussion. 
Numerical computation have been performed with facilities at Supercomputer Center in ISSP, Information Technology Center, 
University of Tokyo, and with the RIKEN Cluster of Clusters (RICC) facility. 
This work has been supported in part 
by Grant-in-Aid for Scientific Research from MEXT Japan 
under the Grant No. 25287096 and 
by RIKEN iTHES Project.


\begin{thebibliography}{}

\bibitem{Randall} J. J. Randall, L. Katz, and R. Ward, 
\href{http://pubs.acs.org/doi/abs/10.1021/ja01559a004}{J. Am. Chem. Soc. \textbf{79}, 266 (1957)}.

\bibitem{Cava} R. J. Cava, B. Batlogg, K. Kiyono, H. Takagi, J. J. Krajewski, W. F. Peck, Jr., L. W. Rupp, Jr., and C. H. Chen, 
\href{http://journals.aps.org/prb/abstract/10.1103/PhysRevB.49.11890}
{Phys. Rev. B {\bf 49}, 11890 (1994)}. 

\bibitem{Shimura} T. Shimura, Y. Inaguma, T. Nakamura, M. Itoh, and Y. Morii, 
\href{http://journals.aps.org/prb/abstract/10.1103/PhysRevB.52.9143} 
{Phys. Rev. B {\bf 52}, 9143 (1995)}. 

\bibitem{Cao}G. Cao, J. Bolivar, S. McCall, J. E. Crow, and R. P. Guertin, 
\href{http://journals.aps.org/prb/abstract/10.1103/PhysRevB.57.R11039}
{Phys. Rev. B {\bf57}, R11039 (1998). }

\bibitem{Okabe} H. Okabe, M. Isobe, E. Takayama-Muromachi, A. Koda, S. Takeshita, M. Hiraishi, M. Miyazaki, R. Kadono, 
Y. Miyake, and J. Akimitsu,
\href{http://journals.aps.org/prb/abstract/10.1103/PhysRevB.83.155118 }
{Phys. Rev. B {\bf 83}, 155118 (2011)}. 

\bibitem{Krempa} W. Witczak-Krempa, G. Chen, Y. B. Kim, and L. Balents, 
\href{http://www.annualreviews.org/doi/abs/10.1146/annurev-conmatphys-020911-125138}
{Annu. Rev. Condens. Mattter Phys. {\bf5}, 57 (2014)}.

\bibitem{Watanabe3} H. Watanabe, T. Shirakawa, and S. Yunoki, 
\href{http://journals.aps.org/prb/abstract/10.1103/PhysRevB.89.165115} 
{Phys. Rev. B {\bf 89}, 165115 (2014)}. 

\bibitem{Sugano} S. Sugano, Y. Tanabe, and H. Kamimura, {\it Multiplets of Transition-Metal Ions in Crystals} 
(Academic Press, New York, 1970). 

\bibitem{Matsuno} J. Matsuno, Y. Okimoto, Z. Fang, X. Z. Yu, Y. Matsui, N. Nagaosa, M. Kawasaki, and Y. Tokura, 
\href{http://journals.aps.org/prl/abstract/10.1103/PhysRevLett.93.167202} 
{Phys. Rev. Lett. {\bf93}, 167202 (2004). }

\bibitem{Wang} X. L. Wang and E. Takayama-Muromachi, 
\href{http://journals.aps.org/prb/abstract/10.1103/PhysRevB.72.064401}
{Phys. Rev. B {\bf 72}, 064401 (2005)}. 

\bibitem{Kim3} B. J. Kim, Jaejun Yu, H. Koh, I. Nagai, S. I. Ikeda, S. -J. Oh, and C. Kim, 
\href{http://journals.aps.org/prl/abstract/10.1103/PhysRevLett.97.106401}
{Phys. Rev. Lett. {\bf97}, 106401 (2006).}

\bibitem{Perry} R. S. Perry, F. Baumberger, L. Balicas, N. Kikugawa, N. J. C. Ingle, A. Rost, J. F. Mercure, Y. Maeno, Z. X. Shen, 
and A P Mackenzie, 
\href{http://iopscience.iop.org/1367-2630/8/9/175} {New J. Phys. {\bf 8}, 175 (2006)}. 

\bibitem{Nagai} I. Nagai, N. Shirakawa, N. Umeyama, and S. Ikeda, 
\href{http://journals.jps.jp/doi/abs/10.1143/JPSJ.79.114719} {J. Phys. Soc. Jpn. {\bf 79}, 114719 (2010)}. 

\bibitem{Kim1} B. J. Kim, Hosub Jin, S. J. Moon, J.-Y. Kim, B.-G. Park, C. S. Leem, Jaejun Yu, T. W. Noh, C. Kim, S.-J. Oh, J.-H. Park, V. Durairaj, G. Cao, and E. Rotenberg, 
\href{http://journals.aps.org/prl/abstract/10.1103/PhysRevLett.101.076402}
{Phys. Rev. Lett. {\bf101}, 076402 (2008).}

\bibitem{Kim2} B. J. Kim, H. Ohsumi, T. Komesu, S. Sakai, T. Morita, H. Takagi, T. Arima, 
\href{http://www.sciencemag.org/content/323/5919/1329.abstract}
{Science {\bf323}, 1329 (2009).}

\bibitem{Ishii} K. Ishii, I. Jarrige, M. Yoshida, K. Ikeuchi, J. Mizuki, K. Ohashi, T. Takayama, J. Matsuno, and H. Takagi, 
\href{http://journals.aps.org/prb/abstract/10.1103/PhysRevB.83.115121}
{Phys. Rev. B {\bf 83}, 115121 (2011)}.

\bibitem{Fujiyama} S. Fujiyama, H. Ohsumi, T. Komesu, J. Matsuno, B.J. Kim, M. Takata, T. Arima, and H. Takagi, 
\href{http://journals.aps.org/prl/abstract/10.1103/PhysRevLett.108.247212}
{Phys. Rev. Lett. {\bf 108}, 247212 (2012). }

\bibitem{Jackeli} G. Jackeli and G. Khaliullin, 
 \href{http://journals.aps.org/prl/abstract/10.1103/PhysRevLett.102.017205} 
{Phys. Rev. Lett. \textbf{102}, 017205 (2009).}

\bibitem{Jin} H. Jin, H. Jeong, T. Ozaki, and J. Yu, 
 \href{http://journals.aps.org/prb/abstract/10.1103/PhysRevB.80.075112} 
{Phys. Rev. B \textbf{80}, 075112 (2009).} 

\bibitem{Watanabe1} H. Watanabe, T. Shirakawa, and S. Yunoki, 
 \href{http://journals.aps.org/prl/abstract/10.1103/PhysRevLett.105.216410} 
{Phys. Rev. Lett. {\bf105}, 216410 (2010).}

\bibitem{Shirakawa} T. Shirakawa, H. Watanabe, and S. Yunoki,
 \href{http://m.iopscience.iop.org/1742-6596/273/1/012148/} 
{J. Phys.:Conf. Ser. {\bf273}, 012148 (2011).}

\bibitem{Martins} C. Martins, M. Aichhorn, L. Vaugier, and S. Biermann, 
 \href{http://journals.aps.org/prl/abstract/10.1103/PhysRevLett.107.266404} 
{Phys. Rev. Lett. {\bf107}, 266404 (2011).}

\bibitem{Arita} R. Arita, J. Kune{\v s}, A. V. Kozhevnikov, A. G. Eguiluz, and M. Imada,
 \href{http://journals.aps.org/prl/abstract/10.1103/PhysRevLett.108.086403} 
 {Phys. Rev. Lett. {\bf108}, 086403 (2012).}
 
\bibitem{Onishi} H. Onishi, 
\href{http://iopscience.iop.org/1742-6596/391/1/012102/} 
{J. Phys.: Conf. Ser. {\bf 391}, 012102 (2012)}. 

\bibitem{Wang2} F. Wang and T. Senthil,
 \href{http://journals.aps.org/prl/abstract/10.1103/PhysRevLett.106.136402}
{Phys. Rev. Lett. {\bf 106}, 136402 (2011).}

\bibitem{Watanabe2} H. Watanabe, T. Shirakawa, and S. Yunoki,
 \href{http://journals.aps.org/prl/abstract/10.1103/PhysRevLett.110.027002}
{Phys. Rev. Lett. {\bf 110}, 027002 (2013). }

\bibitem{Yang} Y. Yang, W.-S. Wang, J.-G. Liu, H. Chen, J.-H. Dai, and Q.-H. Wang, 
\href{http://journals.aps.org/prb/abstract/10.1103/PhysRevB.89.094518} {Phys. Rev. B {\bf 89}, 094518 (2014)}. 
\bibitem{Meng} Z. Y. Meng, Y. B. Kim, and H.-Y. Kee, 
\href{http://journals.aps.org/prl/abstract/10.1103/PhysRevLett.113.177003} 
{Phys. Rev. Lett. {\bf 113}, 177003 (2014). }

\bibitem{MODMFT} G. Kotliar, S. Y. Savrasov, G. P{\'a}lsson, and G. Biroli,
\href{http://journals.aps.org/prl/abstract/10.1103/PhysRevLett.87.186401}
{Phys. Rev. Lett. {\bf87}, 186401 (2001).}

\bibitem{CTQMC} P. Werner, A. Comanac, L. de’ Medici, M. Troyer, and A. J. Millis ,
\href{http://journals.aps.org/prl/abstract/10.1103/PhysRevLett.97.076405}
{Phys. Rev. Lett. {\bf97}, 076405 (2006).}

\bibitem{Eckstein} M. Eckstein, M. Kollar, K. Byczuk, and D. Vollhardt, 
\href{http://journals.aps.org/prb/abstract/10.1103/PhysRevB.71.235119}
{Phys. Rev. B. {\bf71}, 235119 (2005).}

\bibitem{Metzner} W. Metzner and D. Vollhardt, 
\href{http://journals.aps.org/prl/abstract/10.1103/PhysRevLett.62.324}
{Phys. Rev. Lett. {\bf 62}, 324 (1989).}

\bibitem{Kanamori} J. Kanamori, 
\href{}
{Prog. Theor. Phys. \textbf{30}, 275 (1963). }

\bibitem{problem} T. Pruschke and R. Bulla, 
\href{http://link.springer.com/article/10.1140/epjb/e2005-00117-4}
{Eur. Phys. J. B {\bf 44}, 217-224 (2005).}

\bibitem{Sato}  For example, see T. Sato, K. Hattori, and H. Tsunetsugu, 
\href{http://journals.aps.org/prb/abstract/10.1103/PhysRevB.86.235137}
{Phys. Rev. B {\bf 86}, 235137 (2012)}. 

\bibitem{MEM} M. Jarrell and J. E. Gubernatis,
\href{http://www.sciencedirect.com/science/article/pii/0370157395000747}
{Phys. Rep. {\bf 269}, 133 (1996).}

\bibitem{paraMott} A. Georges, S. Florens, and T. A. Costi, 
\href{}
{J. Phys. IV (Colloque) {\bf114}, 165 (2004).}

\bibitem{note2} Without loss of generality, we can consider the magnetic orders along $z$ direction simply because our model is 
isotropic. 

\bibitem{note3} Notice that the off-diagonal elements of $F_{j,m}^{j',m'}(\omega)$ 
correspond to the anomalous single-particle excitations 
for a superconducting state. See, for example, J. R. Schrieffer, {\it Theory of Superconductivity} (Addison-Wesley, New York, 1988). 

\bibitem{ANG} For calculations of the off-diagonal elements, see T. Sato and H. Tsunetsugu,
\href{http://http://journals.aps.org/prb/abstract/10.1103/PhysRevB.90.115114}
{Phys. Rev. B {\bf90}, 115114 (2014). }

\bibitem{note7} 
This implies that the exciton has total angular momentum $J_{\rm EX}=1$ and its $z$ component $M_{J_{\rm EX}}=0$. 

\bibitem{OS-3} C. K. Chan, P. Werner, and A. J. Millis,
\href{http://journals.aps.org/prb/abstract/10.1103/PhysRevB.80.235114}
{Phys. Rev. B {\bf80}, 235114 (2009).}

\bibitem{note4} Notice that $\lambda^*=\lambda$ in the non-interacting limit [see Fig.~\ref{fig:dos-pm}(c)] and 
that $\lambda^*=0$ when $\lambda=0$. 


\bibitem{note5} To examine the effect of a tetragonal crystal field, we have added 
$H_\delta=\delta\sum_{i,\sigma}n_{i\sigma}^{d_{xy}}$ into $H$. In the DMFT calculations, 
we have introduced the magnetic order parameters, 
$M_{\alpha,s}(l)=\frac{1}{2}\sum_{s^{\prime}=\pm s}sign(s^{\prime})\langle b_{l\alpha s^{\prime}}^{\dagger} b_{l\alpha s^{\prime}}\rangle$,
and the excitonic order parameters,
$\Delta_{\alpha,s}^{\alpha',s'}(l)=\langle b_{l\alpha s}^{\dagger} b_{l\alpha's'}\rangle $ with $(\alpha,s)\neq(\alpha',s')$, 
where
$b_{l\alpha s}$ is the eigenstates of $H_{0}+H_\delta$ in the atomic limit. 
These bases are related to the $j$ bases, i.e., 
$b_{l1s}=A_{1}a_{l\frac{1}{2}\frac{s}{2}}-sA_{2}a_{l\frac{3}{2}\frac{s}{2}}$,
$b_{l2s}=sA_{2}a_{l\frac{1}{2}\frac{s}{2}}+A_{1}a_{l\frac{3}{2}\frac{s}{2}}$,
and $b_{l3s}=a_{l\frac{3}{2}\frac{-3s}{2}}$, where $A_1$ and $A_2$ are constant.


\end{thebibliography}
\end{document}